\documentclass[10pt, print,
				amsmath, amssymb,
				aps, pre,
				preprintnumbers, numbers,
				sort&compress, nofootinbib,
				twocolumn, superscriptaddress,
				showpacs, colorlinks,
				linkcolor=blue, citecolor=blue
				]{revtex4-1}

\makeatletter
\renewcommand*{\@fnsymbol}[1]{\ensuremath{\ifcase#1\or \dagger\or
											\ddagger \else\@ctrerr\fi}}
\makeatother

\newcommand{\exclude}[1]{}

\usepackage{graphicx}
\usepackage{amsmath}
\usepackage{amssymb}
\usepackage{bm}
\usepackage{hyperref}
\usepackage{pgf}
\usepackage{caption}
\usepackage{subcaption}
\usepackage{adjustbox}
\usepackage{tikz}
\usepackage{pgfplots}
\usepackage{xcolor}
\usepackage{svg}

\begin{document}

\title{Shape driven confluent rigidity transition in curved biological tissues}
\author{Evan Thomas\thanks{evan.thomas@sickkids.ca}}
\affiliation{\small Program in Developmental and Stem Cell Biology, Research Institute, The Hospital for Sick Children, Toronto, Ontario, Canada}
\author{Sevan Hopyan\thanks{sevan.hopyan@sickkids.ca}}
\affiliation{\small Program in Developmental and Stem Cell Biology, Research Institute, The Hospital for Sick Children, Toronto, Ontario, Canada}
\affiliation{\small Department of Molecular Genetics, University of Toronto, Toronto, Ontario, Canada}
\affiliation{\small Division of Orthopaedic Surgery, The Hospital for Sick Children and University of Toronto, Toronto, Ontario, Canada}


\begin{abstract}

Collective cell motions underlie structure formation during embryonic development.
Tissues exhibit emergent multicellular characteristics such as jamming, rigidity transitions, and glassy dynamics, but there remain questions about how those tissue scale dynamics derive from local cell level properties.
Specifically, there has been little consideration of the interplay between local tissue geometry and cellular properties influencing larger scale tissue behaviours.
Here we consider a simple two dimensional computational vertex model for confluent tissue monolayers, which exhibits a rigidity phase transition controlled by the shape index (ratio of perimeter to square root area) of cells, on surfaces of constant curvature.
We show that the critical point for the rigidity transition is a function of curvature such that positively curved systems are likely to be in a less rigid, more fluid, phase.
Likewise, negatively curved systems (saddles) are likely to be in a more rigid, less fluid, phase.
A phase diagram we generate for the curvature and shape index constitutes a testable prediction from the model.
The curvature dependence is interesting because it suggests a natural explanation for more dynamic tissue remodelling and facile growth in regions of higher surface curvature, without invoking the need for biochemical or other physical differences.
This has potential ramifications for our understanding of morphogenesis of budding and branching structures.

\end{abstract}


\maketitle


Developmental processes require carefully controlled tissue properties, facilitating collective cell motions.\cite{Heisenberg:2013, Lecuit:2013}
Emergent behaviours such as density controlled jamming, confluent rigidity transitions, and glassy dynamics are vitally important to structure formation in embryonic development as well as later processes like wound healing.\cite{Weitz:2011, Manning:2013}
The relationship between these tissue scale behaviours and local cell characteristics, for example cell-cell adhesion and cortical tension, together with cell packings and cell geometry has been the focus of numerous recent studies.\cite{Julicher:2007, Julicher:2010, Manning:2014, Manning:2015, Campas:2021}
There has been less focus however on the interplay between tissue geometry, local cell properties, and emergent collective tissue scale material properties.
Here we study the effect of tissue shape on a well studied vertex model for confluent monolayers.

Computational vertex models offer a toolbox for studying large systems with many cells in a mathematically controllable manner with few system parameters.
For a nice review of vertex models discussing different formulations and use cases see \cite{Salbreux:2017}.
In particular, two dimensional apical vertex models serve as useful toy systems for understanding aspects of confluent tissues in monolayer geometries, such as epithelial sheets.
These models are formed by covering a space with polygons representing cells, where edges represent interfaces between cells and vertices convergences of a few such interfaces.
The vertex positions are the dynamical variables and the system energy is generally given in terms of cell areas and edge lengths.

\begin{figure}
	\includesvg[width=0.98\linewidth]{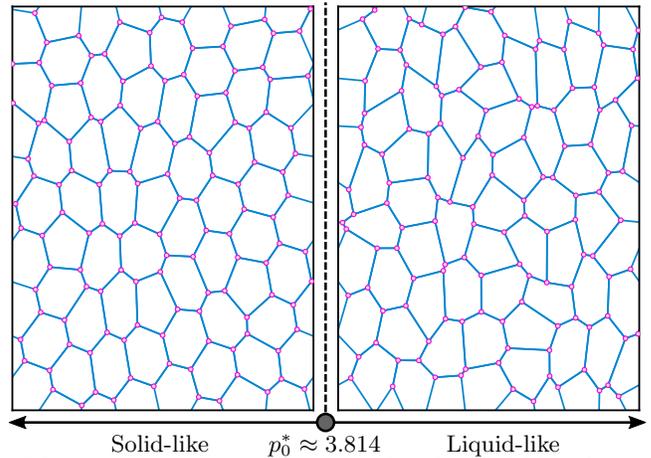}
	\caption{		\label{phase_transition}
The system is more ordered (solid-like) for shape index $p_0$ below the critical value $p_0^* \approx 3.813$ and more disordered (liquid-like) above.
		}
\end{figure}

Some relatively recent work showed that such a simple vertex model (with a standard energy definition) exhibits a phase transition parameterised by the target shape index, the ratio of perimeter of cells to square root of area. \cite{Manning:2014, Manning:2015}
Thus, local cell properties (adhesion, cortical tension), which define the target shape index, determine tissue level material properties (rigidity).
Figure \ref{phase_transition} shows two example systems with target shape indices above and below the critical value.
Even before a careful analysis, it is clear the two systems are qualitatively very different.
The lower target shape index yields a system with more regularly shaped polygons that looks more ordered.
By contrast, the higher target shape index gives a more disordered looking system made up of less regular polygons.

\begin{figure*}
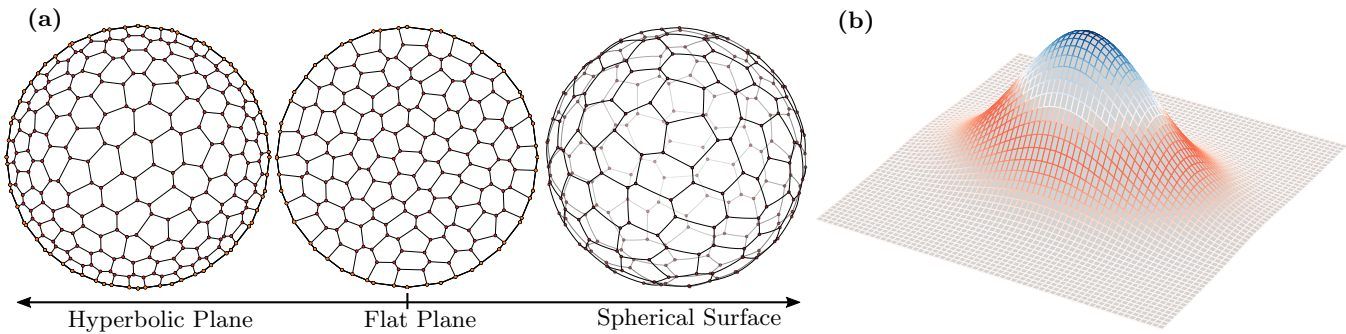

	\begin{subfigure}{0.595\linewidth}
		\begin{tikzpicture}
			\node[inner sep=0pt] (plot) at (0,0)
				{\adjustbox{width=\linewidth}{\centering
					\includesvg[width=\linewidth]{schematic}}};
			\node[align=center] at (-4.8cm,2.1cm) {\bf{(a)}};
		\end{tikzpicture}
		\phantomsubcaption \label{schematic}
	\end{subfigure}
	\begin{subfigure}{0.395\linewidth}
		\begin{tikzpicture}
			\node[inner sep=0pt] (plot) at (-1.5cm,0)
				{\adjustbox{width=\linewidth}{\centering
					\includesvg[width=\linewidth]{bud_rep}}};
			\node[align=center] at (-4.5cm,2.1cm) {\bf{(b)}};
		\end{tikzpicture}
		\phantomsubcaption \label{bud_representation}
	\end{subfigure}
	\caption{		\label{schematics}
(a) Representations of model geometries on the surfaces of interest. The hyperbolic plane has constant negative Gaussian curvature such that each point is a saddle point, and the sphere has constant positive curvature.
(b) Budding type geometry coloured by Gaussian curvature, blue for positive and red for negative. The tip has a positive curvature surrounded by a region of negative curvature toward the base.
		}
\end{figure*}

Phase transitions between more liquid-like easily reconfigurable states and more rigid glassy solid-like states have also been observed in real biological tissues.\cite{Fredberg:2013, Fredberg:2015, Scita:2017, Scita:2019}
Many real tissues, however, have more shape than simply a flat surface.
We know from differential geometry that areas of geodesic polygons on curved surfaces relative to perimeter lengths are adjusted by local Gaussian curvature.
It seems sensible then to consider such vertex models on curved surfaces, and determine what if any effect the surface curvature may have on the phase behaviour, so we have done just that.

The recent paper \cite{Sussman:2020} partially addressed this question in systems of positive Gaussian curvature, but used a different formalism based on the so-called FIRE minimisation scheme \cite{Gumbsch:2006} rather than the Surface Evolver simulations described in \cite{Manning:2015}.
We instead consider both positive and negative curvature in the Surface Evolver context, with positive curvature results confirming \cite{Sussman:2020} and entirely new negative curvature results.

In the following section, we quickly review the relevant parts of the original work \cite{Manning:2015} which first demonstrated the existence of the confluent rigidity phase transition.
This review can also be viewed as an independent verification of their results, as we arrived at effectively the same conclusion by reimplementing their calculations, based on their report, without reference to their original computer code.

Next, we discuss our own results applying similar analysis to nonflat surfaces of constant Gaussian curvature: a sphere for positive and hyperbolic plane for negative curvature.
We carry out the same calculations using the appropriate metric for geodesic lengths and surface areas on the spherical surface embedded in three dimensions or the hyperbolic plane in the Klein disk representation.
Figure \ref{schematic} gives a visual representation of the vertex model system on the different surface geometries, and figure \ref{bud_representation} shows a budding geometry showing where regions of poritive and negative curvature appear.

There does indeed appear to be a curvature dependence for the phase transition critical point.
We find that on more positively curved surfaces there is a lower value of the target shape index at which the system transitions between liquid-like and solid-like behaviour, while on more negatively curved there is a higher value.
In particular, a set of cells with the same properties on a positive curvature surface is more likely to be in a fluid state, with less resistance to configurational remodelling.
Likewise, cells on a negative curvature surface are more likely to be in a solid state.
This curvature dependence constitutes a testable prediction from the computer model.

Finally, we conclude in with an overview of our thoughts about these results, potential ramifications for real world tissues, and the unresolved aspects of this analysis.
Our discussion suggests other modelling or experiments that could support or dispute the validity of these results in physical biological tissues.

The computations described in this paper were carried out primarily with the Surface Evolver software package \cite{Brakke:1992}, which is nicely designed for such calculations.
All of our source code used to run simulations and make plots for this paper are open source and available online.\footnote{
\href{https://github.com/HopyanLab/ConPT2D}{https://github.com/HopyanLab/ConPT2D}}


\section*{Flat Model}									\label{flat_model}

\begin{figure*}
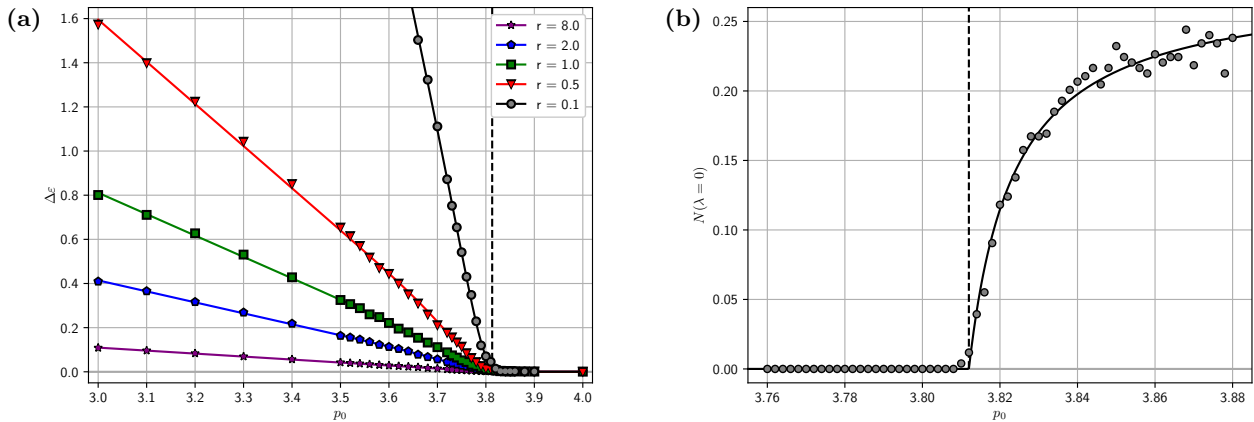

	\centering
	\begin{subfigure}{\columnwidth}
		\begin{tikzpicture}
			\node[inner sep=0pt] (plot) at (0,0)
				{\adjustbox{width=\columnwidth}{\centering
					\input{flat_output_plot.pgf}}};
			\node[align=center] at (-4.1cm,2.3cm) {\bf{(a)}};
		\end{tikzpicture}
		\phantomsubcaption \label{flat_output}
	\end{subfigure}
	\begin{subfigure}{\columnwidth}
		\begin{tikzpicture}
			\node[inner sep=0pt] (plot) at (0,0)
				{\adjustbox{width=\columnwidth}{\centering
					\input{flat_eigen_plot.pgf}}};
			\node[align=center] at (-4.1cm,2.3cm) {\bf{(b)}};
		\end{tikzpicture}
		\phantomsubcaption \label{flat_eigen}
	\end{subfigure}
	\caption{		\label{flat_results}
Flat space plots of (a) mean energy barrier to T1 transitions $\overline{\Delta \varepsilon}$ for a few values of the inverse perimeter modulus $r$ and (b) median density of nontrivial zero modes $\mathcal{N}(\lambda = 0)$ in systems, as a function of target shape index $p_0$. In both plots, the curves show a clear change in behaviour near the dashed line at $p_0 = 3.813$
		}
\end{figure*}

We briefly review the flat surface situation as originally discussed in \cite{Manning:2014, Manning:2015}.
The data presented herein were obtained by repeating the calculations described in \cite{Manning:2015} without reference to any of their simulation code, and can therefore be considered an independent verification of those results.

We consider a two dimensional vertex model for confluent monolayer tissues.\cite{Manning:2014, Manning:2015, Julicher:2007, Shraiman:2007, Carthew:2008, Julicher:2010, Manning:2010, Manning:2012, Shraiman:2012, Shvartsman:2014, Shibata:2015, Sknepnek:2017}
The monolayer is approximated as a collection of polygons tessellating the two dimensional space, here a flat torus so that we do not worry about boundary effects.
The interior of each polygon represents a columnar cell, with the edges being the interfaces between adjacent cells.
The dynamical variables are the positions of the polygon vertices and the energy for each cell is given by
\begin{equation}	\label{original_energy}
	E_i = K \left( A_i - A_0 \right)^2 + \xi P_i^2 + \gamma P_i
\end{equation}

In equation (\ref{original_energy}) the quantities containing the dynamical variables are $A_i$, the cross-sectional area, and $P_i$, the cross-sectional perimeter.
The first term describes resistance to height fluctuations \cite{Shraiman:2007, Angelini:2015, Manning:2015}, with the parameter $K$ denoting the height elasticity and $A_0$ the preferred cross-sectional area for the cell.
The second term describes the active cortical contractility with an elastic constant $\xi$, while the third is an interfacial tension $\gamma$ which models the combination of cortical tension and cell-cell adhesion.\cite{Julicher:2007, Manning:2010}

In principle the model parameters $K$, $A_0$, $\xi$, and $\gamma$ could vary from cell to cell, but we will not consider such complications.
Because a constant term in the energy does not effect the system dynamics, we can complete the square in equation (\ref{original_energy}) to get a nicer form.
Summing over cells, the total system energy is then given by
\begin{equation}	\label{total_energy}
	E = \sum_{i=1}^N \left[ K \left( A_i - A_0 \right)^2 +
							 \xi \left( P_i - P_0 \right)^2 \right]
\end{equation}
where ${P_0 \equiv -\gamma/2\xi}$ is a preferred cross-sectional perimeter for the cells.
With the further definitions ${a_i \equiv A_i/A_0}$, ${p_i \equiv P_i/\sqrt{A_o}}$, and ${\varepsilon \equiv E/KA_0^2}$ the system is described by a nondimensionalised energy
\begin{equation}	\label{energy}
	\varepsilon = \sum_{i=1}^N \left[ \left( a_i - 1 \right)^2 +
							 \frac{1}{r} \left( p_i - p_0 \right)^2 \right]
\end{equation}

In the nondimensionalised energy, equation (\ref{energy}), the dynamical variables are contained in the rescaled areas and perimeters, $a_i$ and $p_i$ respectively, while the two model parameters are the inverse perimeter modulus ${r = K A_0/\xi}$ and the target shape index ${p_0 = P_0/\sqrt{A_0}}$.
As explained in \cite{Manning:2015}, it is sufficient to consider ${p_0 > 0}$, which is equivalent to ${\gamma < 0}$, as the only physically relevant systems.

Within a confluent sheet (absent adding or removing cells) any topological changes must happen through T1 transitions, intercalations wherein an edge separating two neighbours contracts to a point then a new edge forms from the resulting point between two different neighbours.\cite{Weaire:2001, Manning:2014}
Therefore, in order to study tissue properties as a function of the cell properties in this model, we first consider T1 transitions.
Because cells are generally self-propelled, changing topology even in the absence of external forces \cite{Lecuit:2008, Lecuit:2013}, it is natural to consider local resistance to topological rearrangement.

Considering only the ordered ground state, a regular hexagonal tessellation, there is an order-disorder transition at $p_0 = p_0^{hex} \approx 3.722$ \cite{Julicher:2010}, but for higher energy disordered metastable states the situation is different.
As such, in order to study the energy barriers impeding T1 type rearrangements without worrying about boundary effects, we consider ensembles of random metastable states of on a flat torus, a flat Euclidean plane with periodic boundary conditions, for different values of the model parameters $r$ and $p_0$.

For each metastable state we calculated the energy barrier to a T1 neighbour exchange on each edge, forming a distribution of energy barriers as a function of the model parameters.
The mean values of the energy barriers to local T1 rearrangements are plotted in figure \ref{flat_output} as a function of target shape index $p_0$ for a few values of $r$, which reproduces Figure 1b from \cite{Manning:2015}.
As is obvious in figure \ref{flat_output}, there is a sharp change in behaviour around $p_0 = p^*_0 \approx 3.813$, with shape index higher than $p^*_0$ giving systems with very small barriers to local rearrangement.
Below that critical point, energy barriers grow with decreasing shape index with the rate depending on the inverse perimeter modulus, which makes sense as $r$ is just a coefficient on the term containing $p_0$ in equation (\ref{energy}).

\begin{figure*}
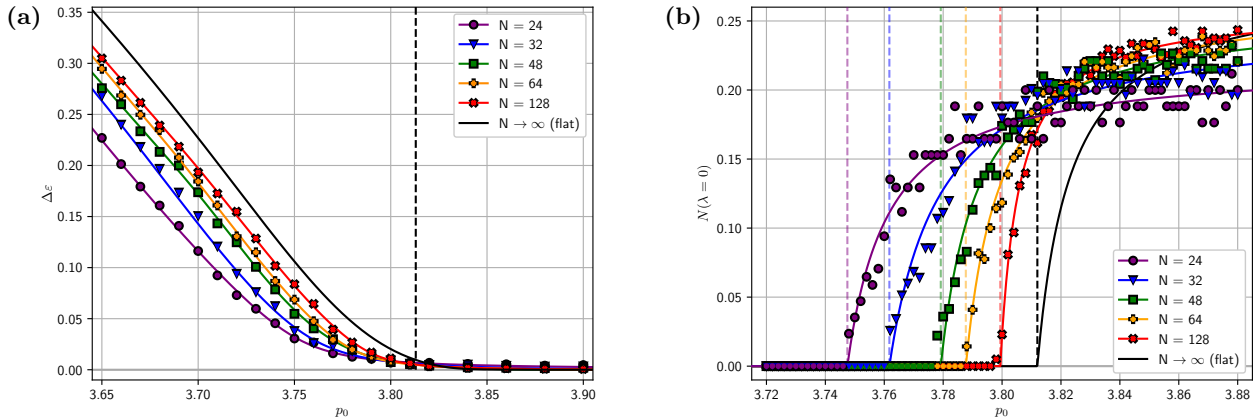

	\centering
	\begin{subfigure}{\columnwidth}
		\begin{tikzpicture}
			\node[inner sep=0pt] (plot) at (0,0)
				{\adjustbox{width=\columnwidth}{
					\input{sphere_output_plot.pgf}}};
			\node[align=center] at (-4.1cm,2.3cm) {\bf{(a)}};
		\end{tikzpicture}
		\phantomsubcaption \label{sphere_output}
	\end{subfigure}
	\begin{subfigure}{\columnwidth}
		\begin{tikzpicture}
			\node[inner sep=0pt] (plot) at (0,0)
				{\adjustbox{width=\columnwidth}{
					\input{sphere_eigen_plot.pgf}}};
			\node[align=center] at (-4.1cm,2.3cm) {\bf{(b)}};
		\end{tikzpicture}
		\phantomsubcaption \label{sphere_eigen}
	\end{subfigure}
	\caption{		\label{sphere_results}
Spherical surface plots of (a) mean energy barrier to T1 transitions $\overline{\Delta \varepsilon}$ as a function of target shape index $p_0$ and (b) median density of nontrivial zero modes $\mathcal{N}(\lambda = 0)$ as a function of target shape index $p_0$ for various sizes of sphere $R = \sqrt{n / 4 \pi }$.
Both families of curves are qualitatively similar to the curves from the flat system in figure \ref{flat_results} but shifted with higher curvature corresponding to lower values of the critical point $p_0^*$.
		}
\end{figure*}

Another way to consider internal rigidity in this system is to look at the spectrum of collective modes rather than barriers to local topological changes.\cite{Ashcroft:1976, Nagel:2005}
Taking the energy given by equation (\ref{energy}) as a function of the positions of the vertices, we form the Hessian matrix
\begin{equation}		\label{hessian}
	H_{ij} = \frac{\partial^2 \varepsilon}{\partial x_i \partial x_j}
\end{equation}
where the $i$ and $j$ indices run over the two spacial dimensions for all the vertices.
The Hessian matrix expresses the curvature of the potential energy at the local minimum defining the metastable state.
Thus, its eigenvalues $\lambda_i$ give the first order approximations to the oscillation frequencies for the normal modes in the system by $\lambda_i = \omega_i^2$.

Furthermore, because the system is at a local minimum, the eigenvalues are positive semidefinite, with $\lambda = 0$ defining the "zero modes" of the system.
Zero modes give the deformations of the system that, at least for small perturbations, have no associated energy cost (eigenvectors with zero eigenvalue).
Denoting the cumulative density of states as $\mathcal{N}(\omega)$, we write the fraction of normal modes in the system which are zero modes (excluding trivial translations) as $\mathcal{N}(\lambda=0)$.

Again assembling ensembles of random metastable states for different parameter choices, the median fraction of zero modes as a function of the shape index parameter is plotted in figure \ref{flat_eigen}.
Notice that, as in figure \ref{flat_output}, the trend in figure \ref{flat_eigen} abruptly changes at $p_0 = p_0^* \approx 3.813 \pm 0.003$, transitioning from no zero modes below that critical point to a significant fraction of the degrees of freedom in the system above (toward $\sim 30\%$).
As in density driven ``jamming'' transitions, for any particular metastable state, the critical point $p_0^i$ is not exactly $p_0^*$, but rather that is the average over the ensemble $p_0^* = \langle p_0^i \rangle$. \cite{Nagel:2002, Manning:2015}

As shown in figure \ref{flat_eigen}, above the critical point the median system acquires "floppy" degrees of freedom, or particular deformations that cost no energy.
This indicates a loss in mechanical stiffness due to the transition to a liquid-like phase.
The appearance of these energetically free collective deformations occurs around the same point that the average energy barrier to local topological reconfigurations approaches zero, as expected.

We have shown the basic elements that are important for our continuation onto curved surfaces, but for a more complete characterisation of this phase transition, including scaling analysis and demonstration that it is similar to a second order phase transition as well as a more thorough discussion of the density of states, see \cite{Manning:2015}.
The next section repeats the calculations of this section on surfaces of constant curvature.


\section*{Curved Model}									\label{curved_model}

\begin{figure*}
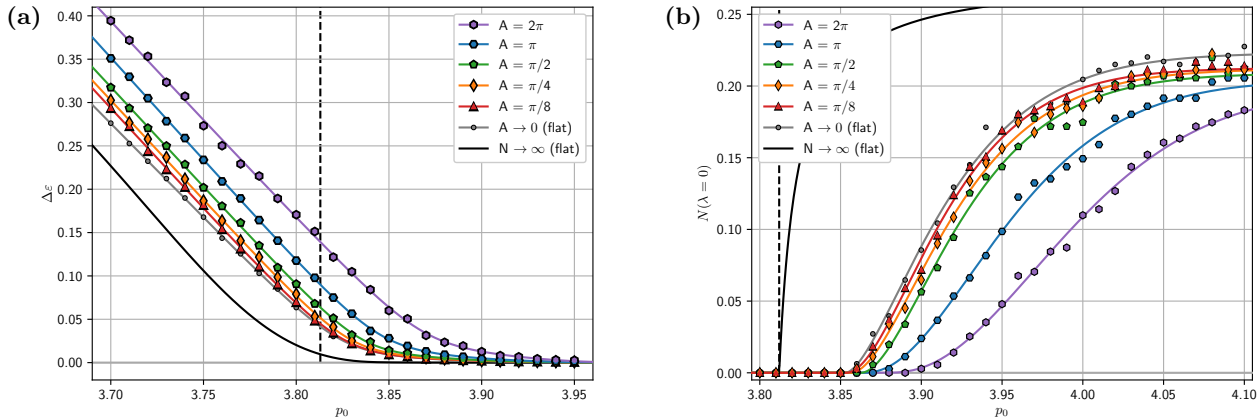

	\centering
	\begin{subfigure}{\columnwidth}
		\begin{tikzpicture}
			\node[inner sep=0pt] (plot) at (0,0)
				{\adjustbox{width=\columnwidth}{
					\input{hyperbolic_output_plot.pgf}}};
			\node[align=center] at (-4.1cm,2.3cm) {\bf{(a)}};
		\end{tikzpicture}
		\phantomsubcaption \label{hyperbolic_output}
	\end{subfigure}
	\begin{subfigure}{\columnwidth}
		\begin{tikzpicture}
			\node[inner sep=0pt] (plot) at (0,0)
				{\adjustbox{width=\columnwidth}{
					\input{hyperbolic_eigen_plot.pgf}}};
			\node[align=center] at (-4.1cm,2.3cm) {\bf{(b)}};
		\end{tikzpicture}
		\phantomsubcaption \label{hyperbolic_eigen}
	\end{subfigure}
	\caption{		\label{hyperbolic_results}
Hyperbolic plane plots of (a) mean energy barrier to T1 transitions $\overline{\Delta \varepsilon}$ as a function of target shape index $p_0$ and (b) median density of nontrivial zero modes $\mathcal{N}(\lambda = 0)$ for a few values of the surface curvature $K = A/96$.
Both families of curves are qualitatively similar to the curves from the flat system in figure \ref{flat_results} but shifted with higher negative curvature corresponding to higher values of the critical point $p_0^*$.
		}
\end{figure*}

In order to address the effect of curvature on the rigidity dynamics, we repeat the analysis from the previous section with systems defined on surfaces of constant Gaussian curvature.
For positive curvature we use a two dimensional spherical surface embedded in three dimensional space.
For negative curvature we use the Klein disk model for the hyperbolic plane.
The energy for the system is still defined by equation (\ref{energy}), but with the appropriate metric used for edge lengths and surface areas.

Because we nondimensionalised the energy with cells having a target area of one, we set the area of the tessellated surface as the total number of cells, $N$.
This is the same as the area of the flat torus in the previous case.
On the sphere we used the entire surface so the curvature of the surface is $K = 1/r^2 = 4\pi/N$.
The hyperbolic plane is infinite in extent, so we used bounded disks of different areas then scaled to the size of our systems.
The curvature of those bounded disks is given by $K = A/N$ where $A$ is the area of the disk chosen before scaling to the size $N$.
In order to compare our results with an experimental system, one would use $KA_c$ in place of our nondimensionalised curvature, where $K$ is the surface curvature and $A_c$ is the average cell cross-sectional area, both measured in the same units of length.

We assembled ensembles of metastable states, as in the flat case, and calculated the energy barriers to all local T1 neighbour exchanges, forming a distribution as a function of the target shape index $p_0$.
The averages (means) of these distributions on various sized spheres are plotted in figure \ref{sphere_output} together with the flat space (black) curve corresponding to the same choice of $r = 0.5$.

The curves in figure \ref{sphere_output} clearly follow the same pattern as the flat situation, very near zero above some critical point $p_0^*$ and finite positive values increasing with decreasing $p_0$ below.
They even have similar slopes in the solid-like regime.
The critical points, however, appear to be staggered with higher curvature (lower $N$) corresponding to lower critical value.

Following the same procedure outlined for the flat model in the previous section, we assembled the Hessian matrix for each spherical system and computed the eigenvalues to find the fraction of (nontrivial) normal modes with zero eigenvalue.
The median fraction for each curvature considered is plotted as a function of the target shape index in figure \ref{sphere_eigen}.

As previously, below a critical shape index there are no nontrivial zero modes in the median system, while above the number increases rapidly toward a sizeable fraction of all modes in the system.
The difference is that the critical point value depends on curvature, following the same pattern as in figure \ref{sphere_output} with higher curvature leading to lower critical value.
The black curves in both are the flat space result from figure \ref{flat_results} for comparison.

Likewise, the negative curvature results for systems on hyperbolic disks of varying curvature are reported in figure \ref{hyperbolic_results}.
On the hyperbolic plane, there is a minor technicality that, because it is non-compact and has nontrivial metric, we had to impose a hard boundary on the disks we chose.
Essentially, the vertices are constrained to stay within the disk.
The imposition of such a boundary serves to both increase the critical point and also soften the transition, such that zero modes appear more gradually in figure \ref{hyperbolic_eigen} than previously in figure \ref{sphere_eigen}.
In order to consistently analyse the results, we repeated the flat space calculations on a flat disk with a boundary.
In figure \ref{hyperbolic_results}, the flat results with a boundary are plotted in grey and labelled $A \rightarrow 0$ in contrast to the original results in black labelled $N \rightarrow \infty$.
The hyperbolic disk results appear to converge toward the bounded flat space results as expected.

The family of curves in figure \ref{hyperbolic_output} again follow the same trend as in flat space, but this time staggered with higher critical value $p_0^*$ corresponding to higher negative curvature (smaller $A$).
The curves in figure \ref{hyperbolic_eigen} also follow that same pattern, but the precise critical value is not as obvious due to the more gradual onset of zero modes in these systems.

\begin{figure}
	\adjustbox{width=\columnwidth}{\input{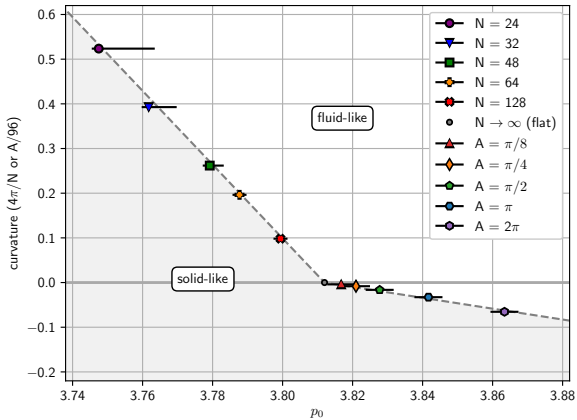}}
	\caption{		\label{phase_diagram}
Phase diagram for systems as a function of curvature ($4\pi/N$ for sphere or $A/96$ for hyperbolic disk) and target shape index $p_0$.
To the left of the curve the system is in a solid-like phase and to the right in a liquid-like phase.
		}
\end{figure}

Estimates for the critical values from both curved space models are plotted in figure \ref{phase_diagram}, giving a rough phase diagram for the confluent vertex model over the parameter space of target shape index and surface curvature.
Parameter choices to the left of the dotted curve will generally describe systems in a more rigid solid-like phase, while those to the right will be a more liquid-like phase.

We should mention that, as \cite{Manning:2015} points out, the distribution of system specific critical values $p_0^i$ is skewed toward lower values for smaller system size.
Then the average critical point $p_0^*$ is also reduced for ensembles of smaller systems, tending toward the thermodynamic limit as the system size considered increases.
The effect is small for the larger systems we considered, but significant for the smaller ones on spheres of higher curvature.
As such, we included the magnitude of this finite size effect, together with the precision of parameters for which we carried out computations, in the error bars reported in figure \ref{phase_diagram}.
The error bars are asymmetric because this effect serves only to reduce the apparent critical value, never increase it.
It is clear also that, although the finite size shift is significant for the smaller systems, it is dwarfed by the adjustment from the curvature dependence.

Given the curvature dependence shown here for the vertex model, we would predict that, in systems of confluent cells making up a sheet type tissue, the tissue would be more fluid and easier to remodel in regions of higher positive Gaussian curvature.
For example, in a budding structure, like a limb bud, we would expect the epithelium to be more dynamic toward the tip where the curvature is higher, and less so nearer the base where the curvature is negative.
We shall comment on this idea more in the next section.

\section*{Conclusion}									\label{conclusion}

We have shown that, in a minimal two dimensional vertex model for confluent monolayer tissues, there is a curvature dependence in the rigidity phase transition.
Specifically, systems with higher positive surface curvature are more likely to be in a liquid-like phase with lower resistance to topological reconfiguration, while those with higher negative surface curvature are more likely to be in a solid-like phase which is harder to remodel.

A curvature dependence of cell rearrangements would have interesting effects on developmental dynamics in physical biological systems.
Various embryonic structures and organ primordia of interest in the study of developmental morphogenesis exhibit highly dynamic cell rearrangements in regions of highest curvature.
Interesting examples include the visceral endoderm of the early mouse embryo,\cite{Srinivas:2012} the optic cup,\cite{Eiraku:2018} the early limb bud.\cite{Hopyan:2015}
Placodes of teeth, salivary, and hair follicles also follow the same pattern,\cite{Breau:2015, Green:2020, Devenport:2018} as do the budding surfaces of branching tubes such as the lung bud, ureteric bud, and mammary gland.\cite{Nelson:2020}

We suggest that, once a budding structure forms, it is dynamically preferable to continue growing outwards since cell neighbour relationships in that section will be easier to rearrange.
In contrast, the saddle curvature at the base of a budding structure would be relatively resistant to cell rearrangements, thereby stabilising and preventing expansion of the base.
In such systems, naturally easier remodelling within the region of comparatively higher curvature would imply a lower reliance on biochemical or mechanical differences to facilitate that dynamical preference.

There remain a few important further issues to address beyond our present scope, however.
We conclude by suggesting some ideas for follow up investigations.

An important follow on would be to ensure our results persist into true three dimensional models.
Because we only considered a two dimensional model, any effect of the complexity of truly three dimensional cell shapes required to cover nonflat surfaces was not incorporated.
For higher curvature, as an example, the tacit assumption of constant cross-sectional area for the cells, used in the vertex model, is increasingly unphysical.
For a nice review of the current state of work generalising vertex models for epithelial tissue see \cite{Escudero:2021}.

Finally, the more important confirmation can come from comparison with experiment.
An experimental investigation would ideally compare rates of T1 neighbour exchanges in a system such as a confluent epithelial sheet over a structure with different curvature regions.
Obvious examples would be compact budding structures like limb buds or branchial arches.
Alternatively, by manipulating the surface curvature either genetically or mechanically it may be possible to adjust tissue rigidity and observe any dynamical changes.


\section*{Methods}

\subsection*{Local Barriers}

We generated random metastable states by making Voronoi diagrams using points distributed by Poisson point process, then relaxing the resulting configuration via gradient descent of energy defined by equation (\ref{energy}) with respect to the dynamical positions of the vertices.
During relaxation, when an edge becomes smaller than a threshold, $l_0 = 0.1$, we allow a T1 neighbour exchange if it reduces system energy.
After such an exchange, in contrast with \cite{Manning:2015}, we separate the two vertices in order to make the resulting edge longer than the threshold.
Lengthening the edges in this way does not seem to change the results, but makes the code execute much more quickly in the liquid-like phase, without getting stuck repeating T1 transitions in a very flat energy landscape.
We were therefore able to set an energy threshold, $\varepsilon_0 = 10^{-8}$, for consecutive minimisation steps and the threshold was generally met in a small number of steps ($\lesssim 1e3$) even in the fluid regime.

After arriving at a suitable metastable configuration, we induce T1 exchanges on each edge in the system and track the total energy change through that process.
In this way, we arrive at a statistical sample of barrier energies to T1 transitions.
In particular, for the flat torus case, we generated $24$ initial Voronoi tessellations of $n = 64$ cells, then relaxed each to a local minimum for different parameter choices.
So, for a particular parameter choice, we arrived at an ensemble of $24$ metastable states.
Each ensemble is generated from the same set of initial Voronoi diagrams.
For a given ensemble of states, we then calculated the system energy as we reduced a given edge down to a point in small steps, which expresses the energy required to force a T1 transition on that edge.
We repeated this process for each edge in each system, so we calculated energy barriers for $3n = 192$ edges in $24$ sample states to get a distribution of $4608$ total simulated energy barriers per parameter choice.
The resulting average (mean) values for each parameter choice are plotted in figure \ref{flat_output}.

In the spherical model, we again assemble ensembles of initial states generated by forming spherical Voronoi diagrams using randomly distributed points on the sphere as centres.
We used the implementation of the spherical Delaunay triangulation presented in \cite{Caroli:2009} included in the SciPy package to generate the Voronoi diagrams.
For all spherical systems the inverse perimeter modulus is chosen as $r = 0.5$.
The choice is arbitrary, except that it controls how quickly the systems "stiffen" below the critical shape index.
As in the flat case, we took each edge and contracted it to a point slowly tracking the system energy to calculate the average barrier to T1 transitions as a function of the shape index parameter, $p_0$.
Here, because we set the curvature by choice of the number of cells that make up the surface, the systems have different numbers of cells by necessity.
We, therefore, choose the number of initial states in the ensemble for each curvature choice by demanding the total number of edge barriers computed be at least that in the flat case discussed in the previous section, $3n\times m \ge 4608$.

In the hyperbolic disk model, we used $96$ cells rather than $64$, since we excluded any edges with vertices touching any cells on the boundary in order to minimise boundary effects to some degree.
Because the number of boundary cells is stochastic and depends on curvature, the number of edges we analysed varried in the range $\sim 100 - 200$.
We therefore used $48$ systems for each parameter choice, so that the total number of edges was again $\ge 4608$ from the flat systems.
We developed our own procedure for randomly distributing points on the disk according to the Klein metric definition of area.
It is included in our source code.

\subsection*{Zero Modes}

To analyse the behaviour of the normal modes in the systems, we again generated ensembles of metastable states in the same manner as discussed for our T1 energy barrier analysis.
Because it is much faster to simply generate metastable states than also calculate energy barriers at all the edges, we increased the size of our ensembles to $96$ states in flat space and $192$ in the curved models.
For each metastable state we exported the Hessian matrix from Surface Evolver at the local minimum, and diagonalised to get the eigenvalues.
(Surface Evolver uses the Hessian for its own relaxation algorithm, so tracks it throughout the simulation.)



\section*{Acknowledgements}

ET would like to thank Ken Brakke and Dapeng Bi for helpful discussions about the workings of Surface Evolver.
This work was funded by the New Frontiers of Research Fund-Exploration (2019-00851) and Canada First Research Excellence Fund/Medicine by Design-New Ideas (2019-03) to SH.


\section*{Author Contributions}

ET and SH designed the project. ET performed the simulations, analysed the data, and wrote the paper. SH supervised the project and edited the paper.


\section*{Competing Interests}

The authors declare no competing interests.


\bibliography{ConPT2D}




\end{document}